\documentclass[aps,prl,twocolumn,showpacs,floatfix]{revtex4}
\usepackage{graphics}
\usepackage{epsfig}
\usepackage{subfigure}
\usepackage{amsmath,amsfonts,amssymb,graphicx}

\begin{document}
\title{Dynamical Coarse Graining of Large Scale-Free Boolean networks}

\author{Wen-Xu Wang$^{1}$}
\author{Gang Yan$^{2}$}
\author{Jie Ren$^{1}$}
\author{Bing-Hong Wang$^{1}$}\email{bhwang@ustc.edu.cn}
\affiliation{ $^{1}$Department of Modern Physics and Nonlinear Science Center,\\
$^{2}$Department of Electronic Science and Technology,
\\
University of Science and Technology of China, Hefei, 230026, PR
China }
\date{\today}

\begin{abstract}
We present a renormalization-grouplike method performed in the
state space for detecting the dynamical behaviors of large
scale-free Boolean networks, especially for the chaotic regime as
well as the edge of chaos. Numerical simulations with different
coarse-graining level show that the state space networks of
scale-free Boolean networks follow universal power-law
distributions of in and out strength, in and out degree, as well
as weight. These interesting results indicate scale-free Boolean
networks still possess self-organized mechanism near the edge of
chaos in the chaotic regime. The number of state nodes as a
function of biased parameter for distinct coarse-graining level
also demonstrates that the power-law behaviors are not the
artifact of coarse-graining procedure. Our work may also shed some
light on the investigation of brain dynamics.
\end{abstract}

\pacs{02.50.Le, 05.65.+b, 87.23.Ge, 87.23.Kg}
\maketitle

Boolean networks are considered to be an important approach for
characterizing the dynamics of complex systems consisting of
interacting units. Examples of such systems cover as diverse as
social and economic networks, neural networks, protein protein
interaction networks and regulatory networks. Random Boolean
network, a very simple and general model, since introduced by
Kauffman in 1969 \cite{RBN1,RBN2}, has drawn much attention from
not only social and biological communities but also physical
community. Intensive investigations indicate that abundant
dynamical properties exist in the random Boolean networks,
classified to ordered and chaotic behaviors \cite{origin}.
However, in recent years, more and more empirical evidences
demonstrate that scale-free (SF) structural properties are
ubiquitous in nature \cite{SF1,SF2}. Therefore, collective Boolean
dynamics on SF networks have been studied for revealing the effect
of structural properties on the dynamics \cite{SFBN}. Dynamics in
the ordered regime of both random and SF networks are fully
explored and attractive cycles (or called attractors) are found,
which is attributed to the self-organized mechanism. However, due
to the restriction of computing ability and the complexity of the
critical dynamical behaviors, a thorough description of the
dynamics in the chaotic regime as well as the edge of chaos,
especially for large networks, remains unclear.

A Boolean network is composed of interacting units (nodes)
${x_1,\cdots, x_n}$. The state of each unit $\sigma(x_i)\in {0,1}
(i=1,\cdots,N)$ is a binary variable. The next time state of any
given unit is determined by both its input from other units or
itself and its assigned Boolean function $F_i$. All the states of
units are allowed to update synchronously. At each time step, the
state of the Boolean network $S(t)$ is denoted by all the $N$
units together: $S(t)=(\sigma(x_1(t)) \ \ \sigma(x_2(t))\ \ \cdots
\ \ \sigma(x_N(t)))$. Thus the state space consists of all the
possible states of the Boolean network. For example, if $N=10000$,
the state space has $2^{10000}$ different states. Each distinct
state $S(t)$ is represented by a node in the state space and
directed links exist between $S(t)$ and $S(t+1)$ with direction
from $S(t)$ to $S(t+1)$. Then the evolution of a Boolean system
can be characterized by a state graph. When the Boolean system is
in the ordered regime, it has been proved that no matter what the
initial state of the system is, the state graph will rapidly
converge to a very small periodic cycle. This behavior is
explained as the ``origin of order" of complex systems. However,
as to the chaotic state as well as the edge of chaos, no one knows
the specific state graph in the state space unless for very small
network size.

In this letter, we present a coarse-graining method to detect the
structure of state graph of SF Boolean networks. Our method is
partially inspired by Kim \cite{Kim}, who proposes a geographic
coarse graining process for detecting the structural properties of
networks of huge size. We first place the Boolean units on the
nodes of a geographical embedded SF network (or called SF network
on lattices) following Ref. \cite{Lattice}. The network is
established as follows: $N=L\times L$ nodes are put on lattice's
points of the two dimensional square net, whereafter the degree
$k$ of each node is chosen according to a given degree
distribution function $P(k)\sim k^{-\alpha}$. Here, we fix
$\alpha=3$ for simplicity. A node $i$ is selected at random and
then its assigned links (the number is $k_i$) are realized on the
basis that the geographically closer vertices are connected first.
Then repeat this procedure until all the nodes are dealt with. As
a remark, this connecting process will lead to a cut off of degree
distribution beyond which $P(k)$ follows a power-law form
$P(k)\sim k^{-\gamma}$. Without losing generality, we also assume
links are bidirectional and each pair of nodes on both sides of
each given link is the input for each other.

After constructing the network structure, we assign each unit
(node) $i=1,\cdots,N$ a Boolean function $F_i$ according to two
``effective" inputs, that is, the state of the unit itself and the
average state of all its neighbors \cite{Input}. The latter is a
majority rule. The input of each unit from its neighbors is
determined by the majority of the neighbors; that is if the
majority's state is $1$, then the input will be $1$; if the number
of units with state $1$ is equal to that with state $0$ among its
neighbors, the input will be $\xi$; otherwise $0$ is inputted.
Therefore, the number of selectable Boolean functions of each unit
is $2^{2 \times 3}=64$ \cite{Input}. The advantage of this
assignment is that the number of the units' input is independent
of their connectivity, which produces certain correspondence to
the classical random Boolean networks. Hence, we can discuss the
effect of the number of input and the  biased parameter $P$
compared with the existent results. Here, $P$ is the probability
of choosing functions with an outcome $0$ and correspondingly with
probability $1-P$ for the functions with an outcome $1$
\cite{SFBN}. For example, $P=1$ means the outcome of the chosen
function for any given input is always $0$; $P=0$ indicates the
output must be $1$. Refer to the Random Boolean networks, the edge
between ordered and chaotic state is described by a function
$2P(1-P)=1$, where $K$ is the average input number of the units
and $P$ is the biased parameter. The case $P=0.5$, $K=2$ is at the
boundary and in this condition the Boolean function space is
composed of $2^{2^K}=16$ different selectable functions. As to our
assignment, the space is composed of $64\simeq 2^{2^{2.6}}$
functions, which indicates that the SF Boolean network is in the
chaotic regime and near the edge of chaos \cite{origin}. Below, we
will provide further evidences for this conclusion.

\begin{figure}
\scalebox{0.70}[0.70]{\includegraphics{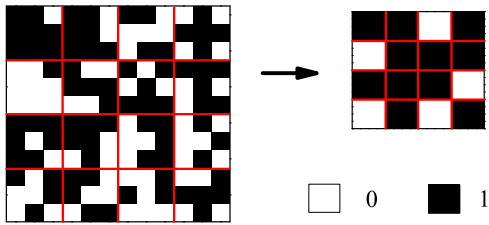}}
\caption{\label{fig:epsart} An example of coarse-graining
procedure. The color coding is that white represents a node with
state $0$ and black a node with state $1$. The original network in
the right side is divided into $4\times 4$ identical size boxes
with $3\times 3$ nodes counted to each box. If the nodes with
state $1$ in a box take the majority, the box's state is $1$;
otherwise it is $0$. Then the system's state is represented by the
coarse-grained graph and the state space is composed of
$2^{4\times 4}$ different states.}
\end{figure}

So far we have given the network structure and the unit
evolutionary rule controlled by the Boolean function, the Boolean
network can evolve step by step and a directed state network is
formed. Unfortunately, when the system is in the chaotic regime,
the network is so huge that no super computers can support it.
Hence, we present a coarse-graining procedure which groups a large
quantities of state nodes at the state space to a single one. This
procedure is described as follows: As shown in Fig. (1), at each
time step, after each unit updates its state, we divide the units
on the square lattice into $m\times m$ square boxes with
\begin{figure}
\scalebox{0.70}[0.70]{\includegraphics{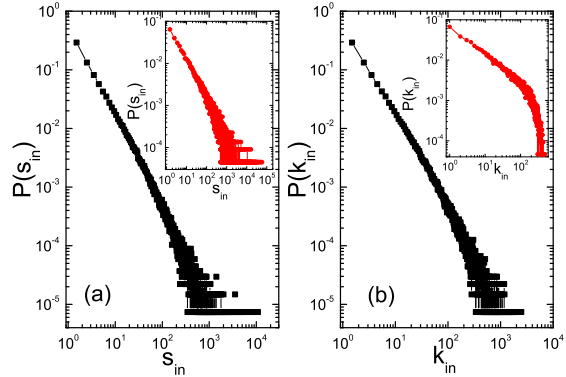}}
\caption{\label{fig:epsart} (a) In-strength and (b) in-degree
distributions of the coarse-grained state network using $4\times
4$ boxes with $N=100\times 100$ Boolean nodes. The insets of (a)
and (b) are the in-strength and in-degree of coarse-grained state
network using $3\times 3$ boxes with $N=99\times 99$ nodes,
respectively.}
\end{figure}
identical size. Each box is considered as a new node whose state
at this step is determined by a majority rule; that is, if the
units with state $1$ in a box take the majority, then the state of
the box is $1$; Otherwise, the box's state is $0$. Therefore the
huge state space of the SF Boolean network is reduced to
$2^{m\times m}$ and the computation ability allows to record every
state and depict the shrank networks in the state space. Note that
this coarse-graining procedure does not affect the evolution of
the Boolean units, but just unbiased shrinks large amount of state
together. Moreover, the statistic properties of the original state
graph can be reflected by the coarse-grained one, which will be
demonstrated later.

\begin{figure}
\scalebox{0.70}[0.70]{\includegraphics{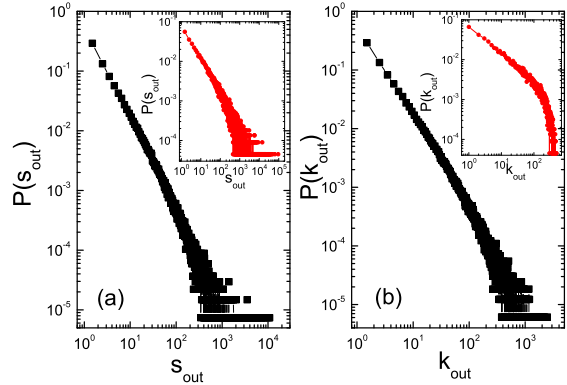}}
\caption{\label{fig:epsart} (a) In-strength and (b) in-degree
distributions of the coarse-grained state network using $4\times
4$ boxes with $N=100\times 100$ Boolean nodes. The insets of (a)
and (b) are results of more heavily coarse-grained state networks
with adopting $3\times 3$ boxes, the original Boolean network size
is $N=99\times 99$.}
\end{figure}

The structural properties of the coarse-grained state networks can
be quantified by the distributions of degree, strength as well as
weight. As mentioned early, the state networks are directed with
each node pointing to the next time state node. Then the basic
structural properties of directed networks including in-degree and
out-degree can be naturally introduced here to characterize the
structural properties of the networks. In-degree of a given node
denotes the number of its neighbors with directed connections
pointing to it. In parallel, out-degree of a node represents the
directed links going out from it. However, the state networks are
far from pure topological structures which will miss important
statistic features. Suppose that a state node arrives at another
node more than one time, this information will be neglected by the
description of node degree. This thus calls for the use of
weighted adjacency matrix element $w_{i\rightarrow j}$
representing the times the state trajectory going from node $i$
immediately to node $j$. A natural generalization of degree in the
case of weighted networks is the node strength (strength for
short). The strength is also divided into in-strength and
our-strength, respectively. The in-strength is defined as
$s^i_{in}=\sum_{j\in \Gamma}w_{j\rightarrow i}$, correspondingly,
the out-strength $s^i_{out}=\sum_{j\in \Gamma}w_{i\rightarrow j}$,
where the sum runs over the neighbor set $\Gamma(i)$ of node $i$.

\begin{figure}
\scalebox{0.70}[0.70]{\includegraphics{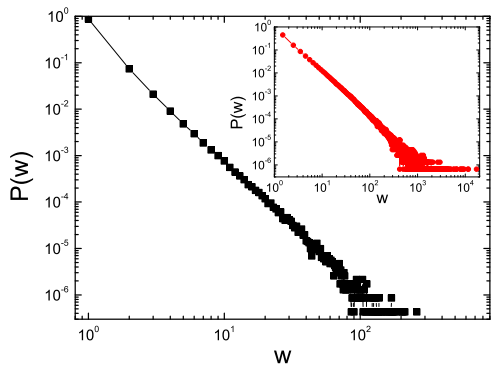}}
\caption{\label{fig:epsart} Weight distribution by adopting
$4\times 4$ boxes and $3\times 3$ boxes (the inset) with original
Boolean network size $N=100\times 100$ and $99\times 99$,
respectively.}
\end{figure}

We now focus on the statistic structural features of the state
networks in terms of the distributions of strength, degree and
weight. We start the simulations from a SF network on lattice with
degree distribution following a power law $P(k)\sim k^{-3}$.
Initial state of each Boolean unit (node) is assigned randomly.
Then each node updates its state based on its randomly chosen
Boolean function from $64$ different functions (we fix the set of
randomly chosen Boolean functions in the course of the time
development. This model is usually called quenched model). After
the updating procedure is finished at each time step, we perform
the coarse-graining procedure and record a state $S(t)$ of the
Boolean network together with a node to denote the state in the
state space. Repeating above procedures for very long time steps,
a weighted network is established. Surprisingly, in the case of
using $4\times 4$ boxes, the coarse-grained network shows perfect
power-law distributions of in-strength and in-degree, out-strength
and out-degree, as well as weight as shown in Fig. 2 to Fig. 4.
The simulations last for $10^6$ time steps. Each distribution is
obtained with an average taken over $20$ different SF network
structures. These results indicate that the state graph in the
chaotic regime near the edge of chaos is highly heterogenous with
few states being reached from or going out to large amount of
other states and most states being reached or going out for few
times. Those highly connected nodes might be judged as attractors.
However, these attractors fundamentally differ from the attractor
cycles existing in the ordered regime. When the state trajectory
falls into an attractor cycle, the complex
\begin{figure}
\scalebox{0.70}[0.70]{\includegraphics{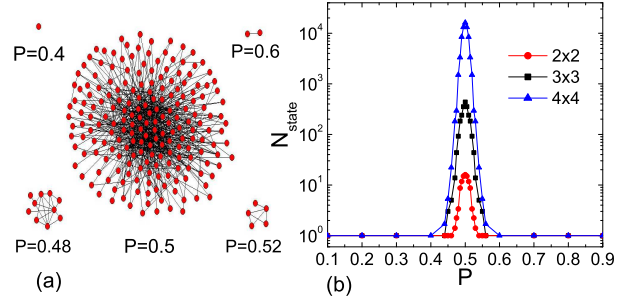}}
\caption{\label{fig:epsart} (a) Prototypical examples of
coarse-grained state networks for different biased parameter $P$.
The simulations only last for $500$ time steps for clear vision.
(b) The number of states in the shrunk state networks by using
$2\times 2$, $3\times 3$ and $4\times 4$ boxes, respectively. A
sharp decline of the state number for different coarse-graining
degree can be observed, which implies a phase transition from
chaotic regime to ordered one.}
\end{figure}
dynamics will determinately evolve along this cycle forever
\cite{origin}. While for the so-called attractors, although the
trajectory falls into an attractor, it still can go out from the
attractor. The distribution of weight also exhibits a power law,
which implies the system evolves from some specific states to
others with high probability. All the five power-law distributions
demonstrate the SF Boolean network experiences a nontrivial
self-organized process in the chaotic regime near the edge of
chaos.

In order to prove that the observed power-law behaviors are not
the artifact of the coarse-graining, we further investigate more
heavily coarse-grained state networks by using $3\times 3$ boxes
with $N=99\times 99$ Boolean nodes, as shown in the insets of Fig.
2 to Fig. 4. It is found that the distributions of strength and
weight display the same power-law distributions in the case of
adopting $4\times 4$ boxes. However, the degree distributions show
the exponential cut off for large degree nodes, which is
attributed to finite size of the state space (for $3\times 3$
boxes, the state space is composed of $2^{3\times 3}=512$
different states). While the power-law distributions of strength
and weight are not influenced by the finite size effect. Moreover,
it is worthwhile to emphasize that all the coarse-grained boxes
contain the same number of Boolean variables, thus each
coarse-grained state node contains the same number of state nodes
in the original state. Suppose that if the original state network
is a random network, then the coarse-grained one must be also a
random network. Overall, we can conclude that the power-law
distributions of the coarse-grained state network is expected to
be the genuine property of the SF Boolean networks, not the
artifact of the coarse-grained information.

To give further evidence for supporting above statement, we
investigate the number of state nodes in the coarse-grained state
network as a function of the biased parameter $P$, as shown in
Fig. 5. One can find that different state nodes for $P=0.5$ are so
many that they nearly fill in the entire coarse-grained state
space. While when the value of $P$ departs from $0.5$, there is a
sharp decline of the number of states, as shown in Fig. 5 (b).
This remarkable change demonstrates a phase transition from
chaotic regime to ordered one. In the cases of $P$ far from $0.5$,
there exists exclusive one node which is the shrunk attractor
cycle in the ordered regime. Therefore, the behaviors of the SF
Boolean network can also be reflected by the coarse-grained state
networks. Fig. 5 (a) shows some examples of state networks for
different value of $P$. The network with $P=0.5$ possesses typical
scale-free features that a few nodes (in the center) have large
amount of connections while most nodes have a few links. When $P$
is slightly larger or lower than $0.5$, few state nodes exist.

\begin{figure}
\scalebox{0.70}[0.70]{\includegraphics{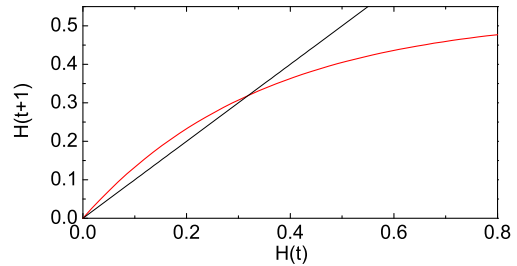}}
\caption{\label{fig:epsart} Fitted Derrida curve of the original
Boolean network for $P=0.5$. The network size is $N=100\times 100$
, $H(t)$ is the Hamming Distance at the time step $t$.}
\end{figure}

We also study the Derrida curve \cite{Derrida} in the case of
$P=0.5$, as exhibited in Fig. 6. The Derrida curve has a nonzero
intersection point with the line of slope $1$. This critical
crossover point (a stable fixed point) demonstrates the system is
indeed in the chaotic regime, as we have mentioned early.

In conclusion, we have investigated the dynamics of large
scale-free Boolean networks. By adopting the coarse-graining
procedure performed in the state space, we find the state networks
near the edge of chaos in the chaotic state exhibit perfect
power-law distributions of in and out strength, in and out degree
as well as weight. The simulations for different coarse-graining
level and different evolutionary time, as well as different biased
parameter demonstrate that the observed universal power-law
distributions are not the artifacts of coarse-graining procedure.
Therefore, we can conclude that the SF Boolean networks in the
chaotic state still perform well-defined self-organized behaviors.

Moreover, since much evidence has suggested that brain i.e. neural
network is in a chaotic state \cite{chaos} and has scale-free
structural properties \cite{brain}, our work in a certain extent
may also be useful for characterizing and understanding the
dynamics of brain, the complexity of which is beyond imagination.
Although the state of each nerve cell can not be measured due to
the experimental limitation, the dynamics of brain still could be
reflected by the coarse-grained measurement.

This work is funded by NNSFC under Grants No. 10472116 and No.
70271070.

\end{document}